\documentclass[aps,prr,showpacs,superscriptaddress,twocolumn,raggedfooter,raggedbottom]{revtex4-2}

\usepackage{graphicx}% Include figure files
\usepackage{dcolumn}% Align table columns on decimal point
\usepackage{bm}% bold math
\usepackage{amsmath,amssymb}
\usepackage{amsthm}
\usepackage{xcolor}

\usepackage{hyperref}% add hypertext capabilities

\begin{document}

\preprint{APS/123-QED}

\title{Quantum arrival times in free fall}

\author{Mathieu Beau}
\affiliation{University of Massachusetts, Boston, MA, USA}
%\affiliation{Massachusetts Institute of Technology, Cambridge, MA, USA}

%\author{Maximilien Barbier}
%\affiliation{Scottish Universities Physics Alliance, University of the West of Scotland, Paisley PA1 2BE, Scotland, United Kingdom}

\author{Timothey Szczepanski}
\affiliation{Ecole Polytechnique, Palaiseau, France}

\author{Rafael Martellini}
\affiliation{Centre International de Valbonne, Sophia-Antipolis, France}

\author{Lionel Martellini}
\affiliation{EDHEC Business School, Nice, France}

\date{\today}% It is always \today, today,

\begin{abstract}

The probability distribution of a time measurement $T_x$  at position $x$ can be inferred from the probability distribution of a position measurement $X_t$ at time $t$ as given by the Born rule \cite{Beau24,Beau24_2}. In an application to free-fall, this finding has been used to predict the existence of a mass-dependent
positive relative shift with respect to the classical time-of-arrival in the long time-of-flight regime for dropped quantum particles \cite{Beau24}. The present paper extends these results in two important directions. We first show that for a Gaussian quantum particle of mass $m$ dropped in a uniform gravitational field $g$, the uncertainties about time and position measurements are related by the relation
$
\Delta T_x \Delta X_t \geq \frac{\hbar}{2mg} .
$
This novel form of uncertainty relation suggests that choosing the initial state so as to obtain a lower uncertainty in the measured position leads to a higher uncertainty in the measured arrival time. Secondly, we examine the case of a free-falling particle starting from a non-Gaussian initial superposed state, for which we predict the presence of gravitationally induced interferences and oscillations in the mean time-of-arrival as a function of the detector's position that can be interpreted as the signature of a Zitterbewegung-like effect.  

\end{abstract}

\maketitle

\pagebreak

\section{Introduction} 

While the Born rule gives the probability distribution of a position measurement at a fixed time, there is no readily available rule in the standard formalism of quantum mechanics for obtaining the probability distribution of a time measurement at a fixed position. This so-called \textit{time-of-arrival (TOA) problem} has been extensively debated in the literature, where a variety of competing approaches have been proposed (see \cite{muga2007time} for a review), but no consensus has emerged so far. The lack of an accepted formalism for the analysis of time-of-arrival can be regarded as one key blind spot in our quantum theoretical description of physical phenomena. In particular, it has been identified as one outstanding difficulty in the formulation of a coherent theory of quantum gravity (\cite{Anderson12} and \cite{Anderson17}). More pragmatically, it is also problematic in the context of interpreting empirical results related to free-falling quantum atoms, which is a prolific area of experimental research. Starting in the nineties, experimental techniques involving cold atoms have been developed to generate empirical TOA distributions using temporal slits (\cite{Dalibard95,Dalibard96,Dalibard96_2}). Recent technological advances have significantly enhanced the precision in the time measurements for free-falling objects, as demonstrated in projects such as MICROSCOPE \cite{Microscope17,Microscope22}, LISA-Pathfinder \cite{LISA18,LISA19}, free-falling matter waves \cite{QuantumTest14}, microgravity experiments on Earth \cite{MicrogravityEarth10,MicrogravityEarth13}, QUANTUS-MAIUS \cite{MicrogravityEarth13,MAIUS18}, and the Bose-Einstein Condensate and Cold Atom Laboratory (BECCAL) \cite{CAL21,CAL22,CAL23}, and it is expected that future experiments, including the Gravitational Behaviour of Anti-hydrogen at Rest (GBAR) experiment \cite{GBAR14,GBAR19,GBAR22,GBAR22_2} and the Space-Time Explorer and Quantum Equivalence Principle Space Test (STE-QUEST) \cite{Altschul15} could lead to further enhancements and yield new experimental insights into the analysis of time in quantum physics (see \cite{ExpReview21} for an extensive review of these developments). 

%In light of the ongoing debates and advancements, further clarification of key concepts in this field is needed, and this is the aim of this paper.
%In light of the ongoing debates and advancements, our paper aims to contribute to the clarification of key concepts in this field.

A particularly striking consequence of our incomplete understanding of the TOA problem is the fact that these empirical results cannot be compared with predictions for the mean and uncertainty of time-of-arrival distributions, which are not available for free-falling particles, or in fact for any quantum system. Some progress in this direction has been made in a recent paper focusing on time measurements for free-falling Gaussian systems with zero initial velocity \cite{Beau24}, where standard results from statistics are used to infer from the Born rule the exact expression for the TOA distribution. The straightforward stochastic representation introduced in \cite{Beau24} can also be used to derive approximate expressions for the mean value and the standard deviation of the TOA of the particle in the semi-classical and in the long time-of-flight (TOF) regime. In the quantum regime where the de Broglie wavelength becomes large compared to the initial spread of the wave packet, one striking finding, which directly follows from a Jensen-type inequality, is that the time of arrival (TOA) of a free-falling quantum particle at a given position is greater than the corresponding classical time-of-arrival, with a delay that depends on the mass of the particle.

In a follow-up article \cite{Beau24_2}, a general method was introduced for calculating the TOA distribution in quantum mechanics for any continuous system, Gaussian or otherwise. This method can be used to predict the probability that a particle will be measured at position $x$ between two instants $t_1$ and $t_2$, and to compute the average TOA and its standard deviation. The approach has been extended to any operator with a continuous spectrum, which leads to the derivation of the the distribution of the time required for a measurement at a given eigenvalue $a$ for an operator $A$. One example of application presented in \cite{Beau24_2} is the distribution of the time of-arrival at a given speed by a quantum particle. Additionally, it is shown in \cite{Beau24_2} that the method is useful for studying the quantum backflow effect, helping to determine the instant at which the current vanishes, thereby providing an experimental signature of the effect.

The present paper extends these results in two important directions. In the Gaussian framework of \cite{Beau24}, we first show that for a quantum particle of mass $m$ dropped in a uniform gravitational field $g$, the uncertainties about time measurements $T_x$ at position $x$ and position measurements $X_t$ at time $t$ are related by the relation
$
\Delta T_x \Delta X_t \geq \frac{\hbar}{2mg},  
$
suggesting that these quantities provide complementary information about the state of the system. Secondly, we use the framework introduced in \cite{Beau24_2} to examine the case of a free-falling particle starting from an initial non-Gaussian state given by a superposition of two Gaussian distributions.  In this case, we predict the presence of gravitationally-induced interferences, reminiscent of a \textit{Zitterbewegung}-like effect, and which imply a non-monotonic relation between free-fall time and the distance to the detector. This effect, which is inherently quantum and particularly pronounced in the near-field regime, could, in principle, lead itself to experimental validation. 

\section{Uncertainty in time-of-arrival measurements for free-falling particles}

In this section, we consider a free-falling particle of mass $m$ in a constant acceleration field $g$ (we consider $g$ to be the acceleration of the gravity field, but it could also be any constant acceleration field caused by other forces, such as an electric field). The Hamiltonian of the particle is given by
\begin{equation}\label{Eq:Hamiltonian}
\widehat{H} = -\frac{\hbar^2}{2m}\frac{\partial^2}{\partial x^2} - mgx\ .
\end{equation}
Notice that if $g>0$, the field is pointing in the same direction as the $x$-axis. 

\subsection{Stochastic representation of the time of arrival}

In what follows we use the stochastic representation recently introduced in \cite{Beau24} to analyze the relation between the random variable, denoted by $T_x$, that describes a time-of-arrival measurement at a given position $x$, and the random variable, denoted by $X_t$, that describes a position measurement at a given time $t$. For the free-falling particle (and also for the free particle, the simple and time-dependent harmonic oscillator, and the constant or time-dependent electric fields), it can be shown that the system stays Gaussian at all times when evolving from a Gaussian initial state (see for example \cite{Klebert73}). 
In this Gaussian setting, and restricting the analysis to a single dimension, the random variable $X_t$ can be written with no loss of generality as \cite{Beau24}:
\begin{equation}\label{Eq:rv:gaussian:t>0}
    X_t  = x_c(t) + \xi\sigma(t),\ 
\end{equation}
where $\xi=\mathcal{N}(0,1)$ is a normally distributed random variable with a variance of $1$ and a mean value of $0$, and where $\sigma(t)$ is the standard-deviation of the Gaussian distribution that is centered at the classical path $x_c(t)$, which by the correspondence principle is also the mean value of the position operator $\langle \hat{x}_t\rangle = \int_{-\infty}^{+\infty}x\rho_t(x)dx = x_c(t)$). By definition and from equation \eqref{Eq:rv:gaussian:t>0}, the random variable $T_x$ is the solution to the equation 
\begin{equation}\label{Eq:DefTx:Gaussian}
x = x_c(T_x) + \xi\sigma(T_x)   
\end{equation}
as the position of the detector is kept fixed at $x$. As outlined in \cite{Beau24}, this equation provides a relation between the random variables $T_x$ and $\xi$, which can be used to derive by the so-called \textit{method of transformation} the distribution of the TOA at a given position $x$ :
\begin{equation}\label{Eq:TOA:Distribution:Gaussian}
    \pi_x(t) =  \left|\frac{v_c(t)\sigma(t)+(x-x_c(t))\dot{\sigma}(t)}{\sigma(t)^2}\right|\times \frac{1}{\sqrt{2\pi}}e^{-\frac{(x-x_c(t))^2}{2\sigma(t)^2}},\
\end{equation}
where $v_c(t)$ is the classical velocity and $\dot{\sigma}(t) = d\sigma(t)/dt$.

The stochastic representation \eqref{Eq:rv:gaussian:t>0} can be used to obtain not only the distribution in \eqref{Eq:TOA:Distribution:Gaussian} but also an explicit characterization of the TOA $T_x$. To see this, we focus on a dropped free falling particle with a zero-mean initial position and a zero-mean initial velocity. In this case, we can write the classical position at time $t$ as $ x_c(t) = \frac{g}{2}t^2$ and the standard deviation of the position at time $t$ as $\sigma(t) = \sqrt{1+\frac{t^2}{\tau^2}}$ with $\tau = \frac{2m\sigma^2}{\hbar}$. Since the distribution of $\xi$ is a normalized Gaussian $\frac{1}{\sqrt{2\pi}}\exp{\left(-\frac{\xi^2}{2}\right)}$, we confirm by a linear transformation that the distribution of the position is $\rho_t(x) = \frac{1}{\sqrt{2\pi \sigma(t)^2}}\exp{\left(-\frac{(x-x_c(t))^2}{2\sigma(t)^2}\right)}$, as expected. From equation \eqref{Eq:DefTx:Gaussian}, we find that the random variable $T_x$ is the solution to the equation
\begin{equation}\label{Eq:FF}
    x = \frac{g}{2}T_x^2 + \sigma \xi \sqrt{1+\frac{T_x^2}{\tau^2}}. 
\end{equation}
This quartic equation can be solved analytically to give the following explicit representation for the TOA $T_x$:
\begin{equation}\label{Eq:TxSol}
    T_x  = t_c \sqrt{1+2q^2\xi^2-2q\xi\sqrt{1+q^2\xi^2+\frac{1}{4q^2}\frac{\sigma^2}{x^2}}},\ \text{with}\ \xi\leq \frac{x}{\sigma}\ ,
\end{equation}
where $t_c =\sqrt{\frac{2x}{g}}$ is the classical time, where the factor $q =  \frac{\hbar}{2m\sigma\sqrt{2gx}} = \frac{\lambda}{4\pi\sigma}$ measures the ratio of the height-dependent de Broglie wavelength ($\lambda=\frac{h}{\sqrt{2mE}}$ with $E=mgx$) to the initial width of the particle wave-packet,
and where the parameter $\beta \equiv \frac{1}{q}\frac{\sigma}{x}$ determines the distinction between the far-field ($\beta \ll 1$) versus near-field ($\beta \gg \max{(1,q)}$) regime of the system.\footnote{The near-field regime is obtained when the term $\beta^2=\frac{1}{4q^2}\frac{\sigma^2}{x^2}$ in equation \eqref{Eq:TxSol} dominates all the other terms, and specifically the term $1+q^2\xi^2$. If $q\xi\ll 1$, then $\beta^2\gg 1$. However, when $\xi^2$ is too large and $q\xi\gg 1$, $\beta^2$ must be substantially larger than that latter term. Since $\xi$ follows a normal distribution with a standard deviation of $1$, it suffices that $\xi$ does not become excessively large compared to $1$ 
%(otherwise, that term becomes negligible due to $e^{-\xi^2/2}\ll 1$). 
Therefore, it is sufficient that $\beta \gg q$ and so that $\frac{\sigma}{x} \gg q^2$. In summary, we conclude that $\frac{\sigma}{x}$ must be significantly larger than the maximum value of $1$ and $q$.}

\subsection{Time/position uncertainty relations}

In what follows, we use \eqref{Eq:TxSol} to obtain a relation between the uncertainty $\Delta T_{x}$ around time measurements and the uncertainty $\Delta X_{t}$ around position measurements:
\begin{equation}\label{Eq:Uncertainty:Farfield}
\Delta T_{x}\Delta X_{t}\geq \frac{\hbar }{2mg}, 
\end{equation}. 
This relation holds for both the near-field and the far-field cases, and within the far-field case both in the semi-classical regime $q\ll 1$ and in the full quantum regime $q\gg 1$. 

\subsubsection{Far-field regime}

To see this, and starting with the far-field regime ($\frac{\sigma}{x} \ll q$), we first note that equation \eqref{Eq:TxSol} can be simplified as:
\begin{align}\label{Eq:TxFarField}
    T_x  &= t_c \left(\sqrt{1+q^2\xi^2}-q\xi\right) \nonumber \\ 
    &\approx 
    \begin{cases}
        t_c\left(1-q\xi + \frac{1}{2}q^2\xi^2\right),\ \text{if}\ q\ll 1 \\
        qt_c\left(|\xi|-\xi\right)\ \ \ \ \  \ \ \ \ \ ,\ \text{if}\ q\gg 1
    \end{cases}\ .
\end{align}
We then compute the uncertainty around the TOA measurement from these expressions. 

For $q\ll 1$, it was shown in \cite{Beau24} (see equation (17) with $v_0=0$) that $\Delta T_x = t_c\frac{\sigma}{\sqrt{2gx}\tau}$, whence 
\begin{equation}\label{Eq:DeltaTOA:farfield:semiclassical}
\Delta T_x = q t_c = \frac{\hbar}{2mg\sigma}, 
\end{equation}
leading to the relation \eqref{Eq:Uncertainty:Farfield} as $\Delta X_t = \sigma(t) = \sigma\sqrt{1+\frac{t^2}{\tau^2}}\geq \sigma$. For $q\gg 1$, we find that 
\begin{equation}\label{Eq:DeltaTOA:nearfield:quantum}
\Delta T_{x}  = qt_c \sqrt{\frac{2(\pi-1)}{\pi}} = \frac{\hbar }{2mg\sigma}\sqrt{\frac{2(\pi-1)}{\pi}} ,
\end{equation}
which leads to
\begin{equation}\label{Eq:Uncertainty:Farfield:qlarge}
\Delta T_{x}\Delta X_{t}\geq \frac{\hbar }{2mg}\sqrt{\frac{2(\pi-1)}{\pi}} , 
\end{equation}
which is greater than $\frac{\hbar }{2mg}$ as $\sqrt{\frac{2(\pi-1)}{\pi}} >1 $.

\subsubsection{Near-field regime}\label{Section:Nearfieldregime}

In the near-field regime ($\frac{\sigma}{x} \gg \max(q,q^2)$), we obtain the following asymptotic relation:
\begin{equation}\label{Eq:TxNearField}
T_x \approx t_c \sqrt{1-\frac{\sigma}{x}\xi},\ \text{with}\ \xi\leq \frac{x}{\sigma}\ ,
\end{equation}
from which we can obtain the approximation for the mean TOA \footnote{Throughout the text, we use the standard notation in statistics $\mathbb{E}(Y)$ to represent the mathematical expectation of a random variable $Y$ (e.g., the TOA $T_x$ or the position $X_t$), while we use instead the notation $\langle\hat{A}\rangle$ to represent the expectation of a quantum operator $\hat{A}$, as is customary in physics.} in the near field regime \cite{SM} 
\begin{equation}\label{Eq:MeanTOA:nearfield}
  \mathbb{E}(T_x)\approx \frac{2^{1/4}\Gamma\left(\frac{3}{4}\right)}{\sqrt{\pi}}t_c\sqrt{\frac{\sigma}{x}} ,
\end{equation}
which has not been derived in \cite{Beau24}. We also have $\mathbb{E}(T_x^2)\approx \sqrt{\frac{2}{\pi}} t_c^2\frac{\sigma}{x}$ \cite{SM}, from which we obtain 
\begin{equation}\label{Eq:DeltaTOA:nearfield}
\Delta T_x \approx k\cdot t_c \sqrt{\frac{\sigma}{x}} = k\cdot \sqrt{\frac{2\sigma}{g}} ,     
\end{equation}
where $k\equiv \sqrt{\sqrt{\frac{2}{\pi}}\left(1-\frac{\Gamma\left(\frac{3}{4}\right)^2}{\sqrt{\pi}}\right)} \approx 0.349$. Finally, we have:
\begin{equation}\label{Eq:Uncertainty:Nearfield}
\Delta T_{x}\Delta X_{t} \geq k\cdot \sqrt{\frac{2\sigma^3}{g}} .
\end{equation}
This relation implies that as the initial uncertainty around the position $\sigma$ becomes very large, the lower bound of \eqref{Eq:Uncertainty:Nearfield} increases as $\sigma^{3/2}$. Since $\sigma$ must at least satisfy the condition $\frac{\sigma}{q x} \gg q$ for the asymptotic relation \eqref{Eq:Uncertainty:Nearfield} to be valid, we find that $\sigma \gg \frac{x_0}{2^{2/3}} \approx 0.630 x_0$, where $x_0 \equiv \left(\frac{\hbar^2}{2m^2g}\right)^{1/3}$ is the characteristic gravitational length \cite{Nesvizhevsky15}, which means that the lower bound in Equation \eqref{Eq:Uncertainty:Nearfield} is very large compared to $\frac{k}{\sqrt{2}}\frac{\hbar}{2mg} \approx 0.247 \frac{\hbar}{2mg}$ and thus very large compared to the lower bound in Equation \eqref{Eq:Uncertainty:Farfield}. In the intermediate regime $\frac{\sigma}{x} \sim \max(q,q^2)$ no analytical expression can be obtained for the lower bound of $\Delta T_x \Delta X_t $, but we confirm the uncertainty relation after a thorough exploration of the parameter space via numerical computation. Taken together, these results suggest that the bound in equation \eqref{Eq:Uncertainty:Farfield} is universal and is valid for all possible values of the parameters $g,\ x,\ m$, and $\sigma$. 

Overall, this new form of uncertainty relation suggests the presence of a mutual exclusiveness between time (of arrival) and position measurements. It is indeed not possible to measure both the time-of-arrival and initial position of particle with an arbitrary precision in the sense that choosing the initial state so as to obtain a lower uncertainty in the measured position inevitably leads to a higher uncertainty in the measured time-of-arrival. 

\subsubsection{Numerical results}

\begin{figure*}
    \centering
    \includegraphics[scale=0.55]{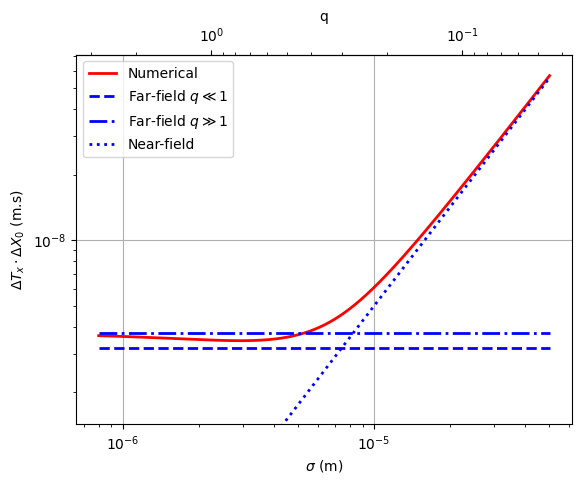}\hspace{0.5cm}\includegraphics[scale=0.55]{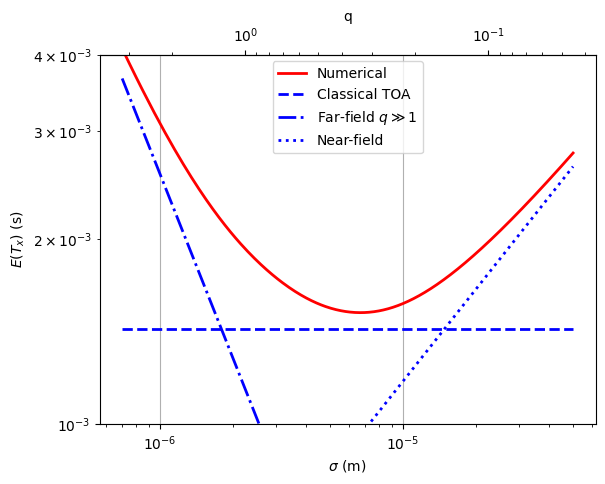}
    \caption{\textbf{TOA-position uncertainty relation and the mean value of TOA for a free-falling particle.} On the left panel, we show the value of the product $\Delta T_x \Delta X_0$ as a function of the initial standard deviation $\Delta X_0 = \sigma$  obtained from the numerical integration of the standard deviation of the stochastic variable given by equation \eqref{Eq:TxSol} (continuous-red-line), as well as the values obtained in the far-field regime as per equation \eqref{Eq:Uncertainty:Farfield} for $q\ll 1$ (dashed-blue-line) and equation \eqref{Eq:Uncertainty:Farfield:qlarge} for $q\gg 1$ (dashed-dotted-blue-line), and in the near-field regime \eqref{Eq:Uncertainty:Nearfield} (dotted-blue-line). As we see from this graph, the bound $\frac{\hbar}{2mg}$ (bottom dashed-blue line) is universal for all regimes. 
    On the right panel, we display the mean values of the TOA (continuous-red-line) as well as the classical TOA $t_c= \sqrt{\frac{2x}{g}}$ (dashed-blue-line) and the two asymptotes corresponding to the near-field (dotted-blue-line) and the far-field with $q\gg 1$ (dashed-dotted-blue-line). In these two graphs, we took the values for $x=10^{-5} \text{m},\ g=9.8 \text{m}\cdot\text{s}^{-2}$ and $m=1.67 \cdot 10^{-27}\text{kg}$ (hydrogen atom). We added a secondary $x-$axis at the top to visualize the evolution of $q$ as a function of $\sigma$ as well.} 
    \label{fig:uncertainty}
\end{figure*}

In Figure \ref{fig:uncertainty}, we report the results of a numerical estimation of the evolution of $\Delta X_0\Delta T_x$ as a function of $10^{-2}\leq\frac{\sigma}{x}\leq 10^1$ for a hydrogen atom of mass $m=1.67\times 10^{-27} \text{kg}$ falling into the Earth's gravitational field $g=9.81 \text{m}\cdot\text{s}^{-2}$ with $x=10^{-5}\text{m}$. Figure \ref{fig:uncertainty} shows that $\frac{\hbar }{2mg}$ is indeed the lowest bound for all regimes, a result that is robust with respect to changes in parameter values. In Figure \ref{fig:uncertainty} (see right panel), we also display the mean values of the TOA as well as the classical TOA $t_c= \sqrt{\frac{2x}{g}}$ and the two asymptotes corresponding to the near-field and the far-field with $q\gg 1$. We confirm through this analysis the finding reported in \cite{Beau24} that the mean value of the TOA is greater than the classical TOA in the far-field regime, and extend this result by showing that it also holds in the near-field regime.

\subsubsection{Time-energy uncertainty relation}

As a side comment, we note that the standard time-energy uncertainty relation can be confirmed within our framework. To see this, we first compute the uncertainty for the energy of a free-falling particle as $\Delta E = mg\sigma\sqrt{1+\frac{\hbar^4}{32g^2m^4\sigma^6}}$ \cite{SM}, and then combine this result with equation \eqref{Eq:Uncertainty:Farfield} to obtain as expected
\begin{equation}\label{Eq:TimeEnergyUncertainty}
    \Delta E \Delta T_x \geq \frac{\hbar}{2} .
\end{equation}
The time-energy uncertainty relation that we confirm here relates to a joint analysis of the dispersion of energy measurements for a quantum system and the dispersion of time-of-arrival measurements at a given position $x$. This interpretation of the time-energy uncertainty relation closely aligns with that presented in \cite{Dalibard96}, where the authors experimentally measure the dispersion in the time-of-arrival and compare it with the dispersion of the energy of the system. This is in contrast with competing interpretations of the time/energy relation, where time is rather regarded as the time of transition between different energy states (see \cite{Busch2008} for an extensive review). 
 
\section{Gravitationally-induced interference for a free-falling particle in a superposed state}

We now examine the case of a free-falling particle starting from a non-Gaussian initial superposed state using the framework introduced in \cite{Beau24_2}.

\begin{figure*}
    \centering
    \includegraphics[width=0.45\linewidth]{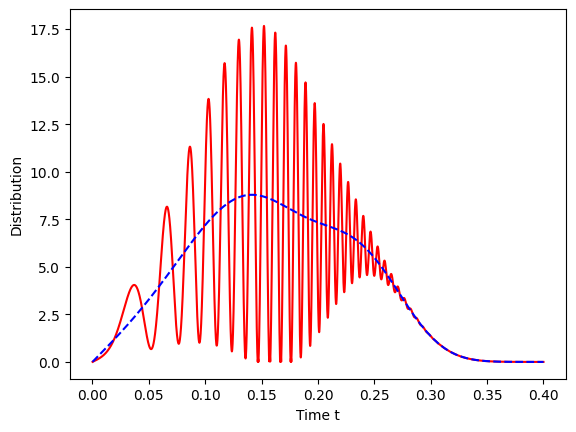}\includegraphics[width=0.44\linewidth]{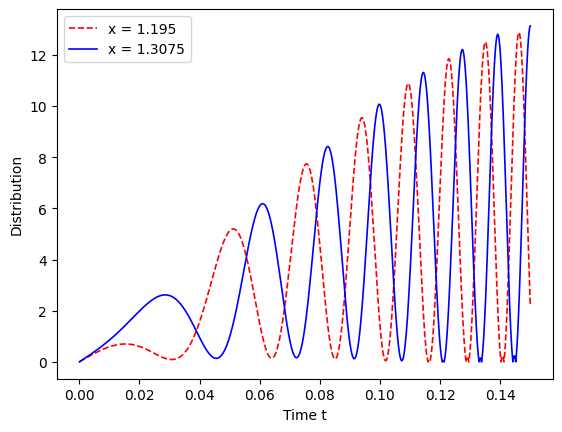}
    \caption{\textbf{TOA distributions for a free-falling superposition of two Gaussian wavepackets.} On the left panel, we plot the TOA distribution (continuous-line) for the free fall (the acceleration of gravity is $g=200$) of a particle of mass $m=1$ (we also take $\hbar=1$) in a superposition of two Gaussian wave packets with an initial width $\sigma = 3$ and centered at $x=0$, with two opposite initial wave vector $k_1=10$ and $k_2=-10$. The detector is located at $x=2$. The dashed-line represents the average TOA distribution for the two Gaussian without interference, contrasting with the gravitational-induced interference shown in the continuous curve. On the right panel, we plot the TOA distribution with the same parameters, except that we take $x=x_1=1.195$ for the continuous-line and $x=x_2=1.3075$ for the dashed line, showing the displacement of the peaks and the distribution's minima slightly to the left for $x=x_2$. }
    \label{Fig:TOAdistrib:Superposition}
\end{figure*}

\begin{figure*}
    \centering
    \includegraphics[width=0.45\linewidth]{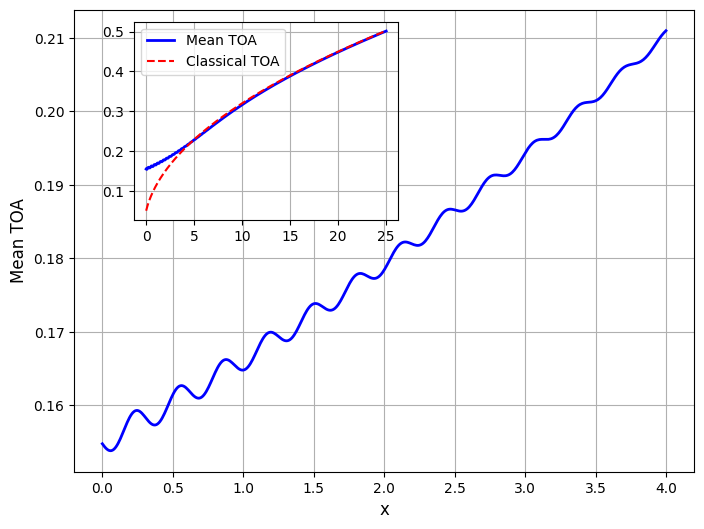}\includegraphics[width=0.455\linewidth]{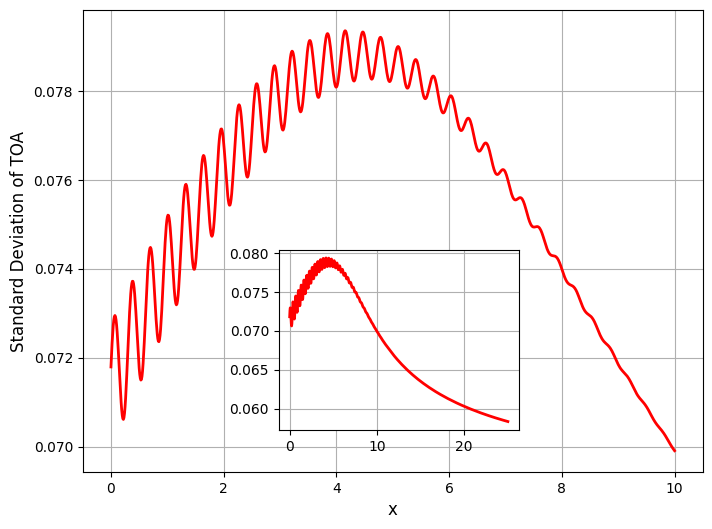}
    \caption{\textbf{Mean and standard deviation of the TOA for a free-falling superposition of two Gaussian wavepackets.} In this figure, we plot the mean TOA vs. the position of the detector $x$ (continuous-line, large panel on the left) and the standard deviation of the TOA vs. $x$ (continuous-line, large panel on the right) for the free fall (the acceleration of gravity is $g=200$) of a particle of mass $m=1$ (we also take $\hbar=1$) in a superposition of two Gaussian wave packets with an initial width $\sigma = 3$ and centered at $x=0$, with two opposite initial wave vector $k_1=10$ and $k_2=-10$. Both panels include sub-panels that display the graphs on a larger scale of $x$, demonstrating a perfect match of the mean TOA with the classical prediction (dashed-line) for large values of $x$ in the left panel, and the decay of the standard deviation for large values of $x$ in the right panel. Notably, the oscillations vanish as $x$ increases.}
    \label{Fig:MeanStDevTOA:Superposition}
\end{figure*}

\begin{figure*}
    \centering
    \includegraphics[width=0.45\linewidth]{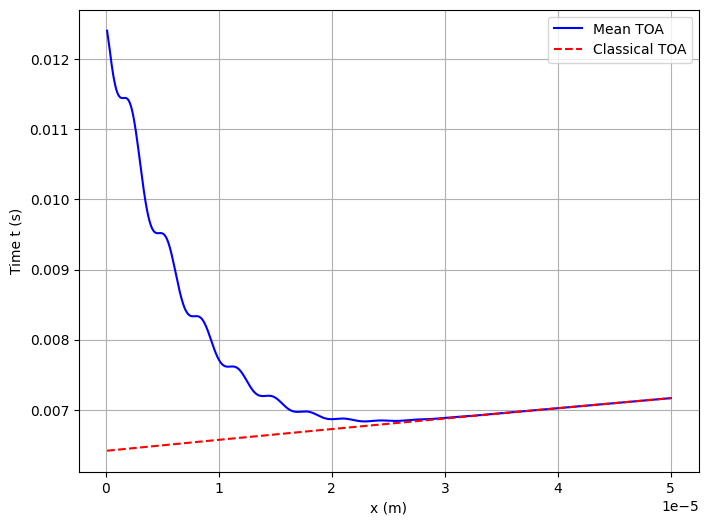}\includegraphics[width=0.455\linewidth]{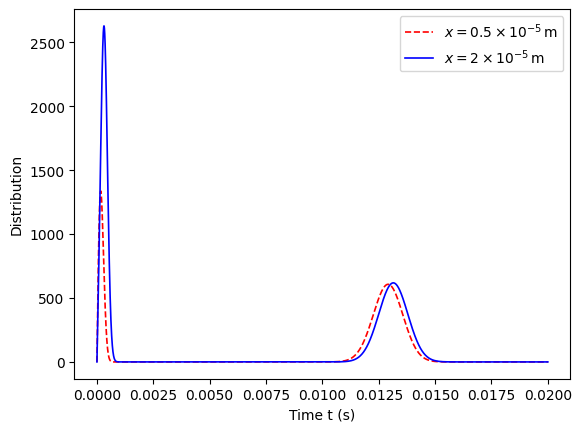}
    \caption{\textbf{Mean TOA for a free-falling hydrogen atom in the superposition of two Gaussian wavepackets.} On the left panel, we show the mean TOA (continuous-line) and the classical TOA (dashed-line) versus the position $x$ of the detector for a hydrogen atom with mass $m = 1.67 \times 10^{-27} \text{ kg}$, initially prepared in a superposition of two Gaussian wave packets of width $10^{-5} \text{ m}$, centered at $x = 0$, with respective wave vectors $k_1 = -k_2 = 10^{-6} \text{ m}^{-1}$, freely falling in Earth's gravitational field $g = 9.8 \text{ m} \cdot \text{s}^{-2}$. On the right panel, we plot two TOA distributions as a function of $t$ for the same system with $x_1 = 0.5 \times 10^{-5} \text{ m}$ (dashed-line) and $x_2 = 2 \times 10^{-5} \text{ m}$ (continuous-line).} 
    \label{Fig:MeanSt:Superposition:Hydrogen}
\end{figure*}

\subsection{Time of arrival distributions for non-Gaussian quantum systems}

It is shown in \cite{Beau24_2} that for any continuous quantum system (Gaussian or otherwise) and any observable $A_{t}$, the distribution $ \pi _{a}\left( t\right) $ of a time measurement at a fixed state $a$ can be inferred from the distribution $ \rho _{t}\left( a\right) $ of a state measurement at a fixed time $t$ via the transformation $ \pi _{a}\left( t\right) \propto \left\vert \frac{\partial }{\partial t} \int_{-\infty }^{a}\rho _{t}\left( u\right) du \right\vert $. This analysis generalizes to non-Gaussian distributions the framework introduced in \cite{Beau24} for Gaussian systems. It can be used to relate the distribution of the time-of-arrival to a given position to the absolute value of the probability current at that position. More specifically, the probability
distribution function, denoted by $\pi
_{x}\left( t\right) $, of the measured TOA $T_{x}$ is given by (see Proposition 2 in \cite{Beau24_2}):%
\begin{equation}
\pi _{x}\left( t\right) =\mathcal{K} \left\vert \frac{\partial }{\partial t}F_{t}\left(
x\right) \right\vert \text{ } = \mathcal{K}\left\vert
j_{t}\left( x\right) \right\vert,
\label{pi_def}
\end{equation}%
where $ F_{t}\left(x\right) \equiv \int_{-\infty }^{x}\rho _{t}\left( u\right) du $ is the cumulative probability distribution of a position measurement and $j_{t}\left( x\right) \equiv \frac{\hbar }{2mi}\left( \psi ^{\ast
}(t,x)\frac{\partial }{\partial x}\psi (t,x)-\psi (t,x)\frac{\partial }{%
\partial x}\psi ^{\ast }(t,x)\right) $ is the \textit{probability current}, and where $\mathcal{K}$ is a normalization factor.

As an example of application, \cite{Beau24_2} consider the time-of-arrival distribution for a free particle evolving in one dimension with a non-Gaussian initial wave
function given by the superposition of two Gaussian wave packets with
initial position $0$ and wave vectors $k$ and $-k$, respectively. In the next section, we extend the analysis to a free-falling particle in an initial superposed state. 

\subsection{Application to a free-falling  particle in a superposed state}

\subsubsection{Derivation of the current for a superposition of Gaussian states}

In what follows, we use the formalism introduced in \cite{Beau24_2} to generate predictions about TOA distributions for a free-falling particle evolving in one dimension with an initial wave
function which is given by the superposition of two Gaussian wave packets with mean
initial position $0$ and wave vectors $k_1$ and $k_2$. The wave function at time $t$ can be decomposed as 
\begin{equation}\label{Eq:QSuperposition}
\psi\left( x,t\right) =\mathcal{N}\left[ \psi_{1}\left( x,t\right) +\psi_{2}\left( x,t\right) \right] ,
\end{equation}
where $\mathcal{N}$ is a normalization factor and where the wave functions $\psi_{j}$ for $j=1,2$ are given by
\begin{widetext}
\begin{equation}
\psi_{j}\left( x,t\right)  =\frac{1}{\left( 2\pi \sigma(t)^{2}\right)
^{1/4}}e^{-\frac{\left( x-x_{j}(t)\right) ^{2}}{4\sigma_{t}^{2}}} e^{+i\frac{
\lambda(t)^{2}}{8\sigma ^{2}\sigma(t)^{2}}\left( x-x_{j}(t)\right)
^{2}}
e^{i\left[k_{j}(t)x-\frac{\hbar}{2m}k_j^2t-\frac{g}{2}k_jt^2-i\frac{m^2g^2 t^3}{6\hbar}\right]} , \label{Eq:wavefuncgrav}
\end{equation}
\end{widetext}
with $\lambda(t)=\sqrt{\frac{\hbar t}{m}}$, and where the time-dependent wave-vectors are $k_{j}(t)=k_j +\frac{mg}{\hbar}t $, the classical position are $x_{j}(t)=\frac{\hbar k_j}{m}t + \frac{g t^2}{2} $, and where the standard deviation of the position operator $\sigma(t)=\sigma \sqrt{1+\frac{t^{2}}{\tau ^{2}}}$ with $
\tau =\frac{2m\sigma ^{2}}{\hbar }$.

After rewriting the expression of the wave functions \eqref{Eq:wavefuncgrav} in the form 
$$\psi_j\left(x,t\right) = e^{\phi_j}e^{i\varphi_j} ,$$ 
where $\phi_j$ is the real phase and $\varphi_j$ the imaginary phase of $\eqref{Eq:wavefuncgrav}$,
and from the definition of the current (see equation (30) in \cite{Beau24_2}) we find a general expression for the current of the superposition \eqref{Eq:QSuperposition} \footnote{Note that the case of a Gaussian particle with a non-zero initial velocity, which generalizes the model with $v_0=0$ discussed in \cite{Beau24}, can be recovered as a specific case of this analysis by taking $k_1=k_2=k$.}:
\begin{widetext}
\begin{eqnarray}\label{Eq:CurrentGeneral}
j = \mathcal{N}\left[v_1\rho_1 + v_2\rho_2 + (u_1-u_2)\sqrt{\rho_1\rho_2}\sin(\varphi_1-\varphi_2)+(v_1+v_2)\sqrt{\rho_1\rho_2}\cos(\varphi_1-\varphi_2)\right] ,
\end{eqnarray}
\end{widetext}
where:
\begin{align}
    \phi_j &= -\frac{1}{4}\ln\left(2\pi \sigma_j(t)^2\right)-\frac{(x-x_j(t))^2}{4\sigma_t^2}, \label{Eq:phij} \\
    \varphi_j &= \frac{\lambda(t)^2}{8\sigma^2 \sigma(t)^2}(x-x_j(t))^2 + k_j(t)x \nonumber \\ 
    & \ \ \ \ \  \ \ \ \ \  \ \ \  -\frac{\hbar}{2m}k_j^2t-\frac{g}{2}k_jt^2 - \frac{m_j^2 g^2 t^3}{6\hbar}. \label{Eq:varphij} 
\end{align}
and
\begin{align}
      \rho_j &\equiv |\psi_j|^2 = e^{2\phi_j} =\frac{1}{\sqrt{2\pi \sigma_j(t)^2}}e^{-\frac{(x-x_j(t))^2}{2\sigma_t^2}} \\
    u_j &\equiv  \frac{\hbar}{m}\frac{\partial}{\partial x}\phi_j = -\frac{\hbar}{m}\frac{(x-x_j(t))}{2\sigma_t^2} \\
    v_j &\equiv \frac{\hbar}{m}\frac{\partial}{\partial x}\varphi_j = \frac{\hbar}{m}\frac{\lambda(t)^2}{4\sigma^2 \sigma(t)^2}(x-x_j(t)) + \frac{\hbar}{m_j}k_j(t) 
\end{align}

\subsubsection{Numerical results and Zitterbewegung-like effect}

In Fig.~\ref{Fig:TOAdistrib:Superposition}, we present the time-of-arrival (TOA) distribution obtained by plugging equation \eqref{Eq:CurrentGeneral} into \eqref{pi_def} for a particle in free fall under gravity ($g = 200$), with mass $m = 1$ and $\hbar = 1$, initially prepared in a superposition of two Gaussian wave packets of width $\sigma = 3$, centered at $x = 0$, and with opposite initial wave vectors $k_1 = 10$ and $k_2 = -10$. The continuous line represents the TOA distribution at a detector located at $x = 2$, while the dashed line shows the average TOA distribution obtained by considering the two Gaussian components separately, thereby highlighting the interference effects introduced by the gravitational field. To further analyze these effects, Fig.~\ref{Fig:MeanStDevTOA:Superposition} depicts the mean TOA (left panel) and its standard deviation (right panel) as functions of the detector position $x$. The mean TOA (continuous line) follows the classical prediction (dashed line) for large values of $x$, confirming the expected semiclassical behavior. Similarly, the standard deviation of the TOA decreases for large values of $x$, with oscillatory features that vanish at larger distances. The insets in both panels extend the analysis to a broader range of $x$ values, further verifying the classical limit and the suppression of quantum fluctuations at long distances. The oscillation of the mean TOA vs. the distance $x$ observed in Fig. \ref{Fig:MeanStDevTOA:Superposition} are explained by the interference pattern of the TOA distribution shown in \ref{Fig:TOAdistrib:Superposition}. 

One can also observe in Figure \ref{Fig:MeanStDevTOA:Superposition} that sometimes the mean TOA is greater for a detector positioned at $x_2$ (further away) than at a closer position $x_1<x_2$. For example, for $x_1 = 1.195$ and $x_2=1.3075$, we find the respective mean TOA to be $t_1=0.16851$ and $t_2=0.16779$, which gives a relative difference of $0.4 \%$. This outcome may seem counterintuitive from a classical perspective as we expect the mean TOA for $x_2$ to be greater than the one for $x_1$. This can be explained by noticing that the peaks in the interference pattern on the TOA distribution are shifted slightly to the left for $x=x_2$, see Figure on the right panel in \ref{Fig:TOAdistrib:Superposition}, and thus, the probability for having lower TOA, calculated as the area below the curve, is slightly more important for $x=x_2$ than for $x=x_1$, which results in a slightly lower mean TOA for $x=x_2$.   \\

The phenomenon of non-monotonic free-fall time is more pronounced in the near-field regime (where $x \ll q\sigma$, as described in Sec.~\ref{Section:Nearfieldregime}) and becomes more significant for higher initial speeds. Consider, for example, the more realistic case of a hydrogen atom with mass $m = 1.67 \times 10^{-27} \text{ kg}$, initially prepared in a superposition of two Gaussian wave packets of width $10^{-5} \text{ m}$, centered at $x = 0$, with respective wave vectors $k_1 = -k_2 = 10^{6} \text{ m}^{-1}$ (corresponding to a velocity of approximately $6.3 \text{ cm}\cdot\text{s}^{-1}$), freely falling in Earth's gravitational field $g = 9.8 \text{ m} \cdot \text{s}^{-2}$, see Figure \ref{Fig:MeanSt:Superposition:Hydrogen}. In this scenario, if the detector is positioned at $x_1 = 0.5 \times 10^{-5} \text{ m}$ (half the initial width of the wave packets), the mean TOA is $t_1 = 9.52 \times 10^{-3} \text{ s}$. However, if the detector is placed at $x_2 = 2 \times 10^{-5} \text{ m}$, the mean TOA is reduced to $t_2 = 6.87 \times 10^{-3} \text{ s}$, representing a significant $27.8\%$ decrease. This effect arises because the TOA distributions exhibit two distinct peaks, one at $t_{\text{low}} \sim 3 \times 10^{-4} \text{ s}$ and another at $t_{\text{large}} \sim 1.3 \times 10^{-2} \text{ s}$, as can be seen from the right panel in Figure \ref{Fig:MeanSt:Superposition:Hydrogen}. The fact that the amplitude of the TOA distribution peak at $t_{\text{low}}$ is higher and wider when the detector is positioned further away at $x = x_2$ than when it is placed closer to the source at $x = x_1$ explains why the mean TOA is lower when the detector is positioned at $x = x_2$.\\

Interestingly, the oscillations in the mean TOA vs. detector position $x$, as shown in Fig. \ref{Fig:MeanStDevTOA:Superposition}, can be understood by analogy with the \textit{Zitterbewegung effect} in relativistic quantum mechanics, which originates from the interference between positive- and negative-energy solutions of the Dirac equation \cite{FESHBACH58}. Initially predicted by Schrödinger \cite{Schrodinger1930} in the context of the free Dirac equation, the Zitterbewegung effect manifests istelf as rapid oscillations in the position of a relativistic particle due to the coupling between the upper and lower components of the Dirac spinor. In a condensed matter setting, Schliemann et al. \cite{Schliemann05} demonstrated that similar oscillatory motion appears in semiconductor quantum wells due to spin-orbit coupling, thereby providing a non-relativistic analog of this effect. The phenomenon has also been studied in the context of trapped-ion simulations, where Lamata et al. \cite{Lamata07} showed how the Dirac equation and associated relativistic effects, including Zitterbewegung, can be implemented and observed in a controlled quantum system. More recently, Pedernales et al. \cite{Pedernales18} explored the persistence of Zitterbewegung in curved spacetime, establishing a connection between quantum optics and relativistic quantum mechanics in a gravitational field. In our case, the observed oscillations in the mean TOA vs. position of the detector arise from the interference between two Gaussian wave packets, drawing an analogy with the interference between positive- and negative-energy solutions in the relativistic case. 
However, if one calculates the mean value of the position operator as a function of the time $t$ measured in the laboratory, the oscillations disappear, and the expected classical values are recovered. This demonstrates that the two perspectives —mean TOA versus detector position and mean position versus the time at which the detector is active— are not simply the inverse of each other. As a result, experimental measurements of arrival times can reveal curious and unexpected phenomena such as the TOA-Zitterbewegung-like effect.

\section{Outlooks}

In conclusion, this article provides new results regarding time-of-arrival distributions for free-falling quantum particles. From an explicit characterization of the time-of-arrival distribution for Gaussian particles dropped in a constant gravitational field, we first derive a new time/position uncertainty relation that suggests that measurements of time-of-arrival and position are mutually exclusive for a quantum particle. From the experimental perspective, we expect this uncertainty relation to have consequences in quantum metrology, in particular in measures of the acceleration due to gravity based on a detector for which the position is known, as is, for example, the case with the GBAR experiment \cite{GBAR14,GBAR19,GBAR22,GBAR22_2}. 
First, if the position-detector were located close to the source (near-field), then the analysis of the trajectories ought to be performed in the light of the results presented in this paper, in particular equations \eqref{Eq:TxNearField}-\eqref{Eq:Uncertainty:Nearfield} (as well as equations in the paragraph in between) and Figure \ref{fig:uncertainty}. 
Secondly, the uncertainty relation \eqref{Eq:Uncertainty:Farfield} may have an impact on the parameter estimation error of $g$, which would lead to a modification of the Cramer–Rao bound used in \cite{GBAR22}. While beyond the scope of this paper, a modification of the Cramer–Rao bound and its quantum version using a stochastic representation of time-of-arrival would be a possibly relevant avenue for future research. Another potentially interesting metrological application would relate to enhancements of statistical procedures used to determine a detector's position $x$. From experimental measurements of arrival times, we can indeed obtain the position from the empirical estimate for the mean TOA, which explicitly depends on $x$. There again the existence of the time/position uncertainty relation \eqref{Eq:Uncertainty:Farfield} (see also Figure \ref{fig:uncertainty}) implies the presence of intrinsic limitations to the precision of measurements of the detector position $x$ regardless of how small or large will be chosen the initial dispersion value $\sigma$. While our analysis has exclusively focused on free-fall, it would also be interesting to explore whether a time/position uncertainty relation also holds for other systems, such as particles moving in time-dependent electric fields in time-dependent harmonic traps (which could potentially be realized in laboratory settings). These time/position uncertainty relations could be explored for a large variety of quantum systems, both experimentally in laboratories and mathematically within the standard formalism of quantum mechanics. Indeed, the stochastic representation introduced in \cite{Beau24_2} and \cite{Beau24}, further developed in this article, provides a tool for analyzing the time-of-arrival distribution in various quantum systems. 
 
Our analysis leads to several additional new empirical predictions regarding free-falling quantum particles. First, our results suggest that the near-field regime, where the detector is located at a small distance $x$ of the source, and the full quantum regime, where both $x$ and $g$ are extremely small (the latter could be realized in space-based experiments), are the experimental situations where true quantum effects are expected to manifest with greater force. Our results also suggest that it would be of interest to launch experiments with free falling atoms starting from a superposed state, where the oscillatory nature of the predicted Zitterbewegung-like effect should be empirically testable, especially in the near-field regime. It should be noted that a pair of entangled atoms dropped in a gravitational field from a Gaussian state can be analyzed with the same mathematical formalism and would, therefore, represent an alternative experimental setup.

\bibliography{Beau}

\newpage
\onecolumngrid

\vspace{5mm} %5mm vertical space

\begin{center}
\textbf{\large SUPPLEMENTARY MATERIAL}
\end{center}

\vspace{5mm} %5mm vertical space

\section*{Mean value and standard deviation in the near-field regime}

We start with equation \eqref{Eq:TxNearField} in the main body of the article:
$$
T_x \approx t_c \sqrt{1-\frac{\sigma}{x}\xi},\ \xi\leq \frac{x}{\sigma} ,
$$
where $\frac{\sigma}{x}\gg \max(q,q^2)$. We must have $\frac{\sigma}{x} \gg 1$ or else $T_x \approx t_c$, which, in this case, is equivalent to the semi-classical case that we already know. In this regime, $\frac{x}{\sigma} \ll 1$, and thus, we consider only the region where $\xi \leq 0$. Therefore, we have
$$
T_x \approx t_c \sqrt{\frac{\sigma}{x}}\sqrt{|\xi|},\ \xi\leq 0 ,
$$

Let us recall that the mean value of the stochastic variable $T_x$ is given by
$$
\mathbb{E}(T_x) = \frac{1}{\mathcal{N}}\int_{-\infty}^{0}T_x(\xi) \frac{1}{\sqrt{2\pi}}e^{-\frac{\xi^2}{2}} ,
$$
where 
$$
\mathcal{N} = \int_{-\infty}^{0}\frac{1}{\sqrt{2\pi}}e^{-\frac{\xi^2}{2}} = \frac{1}{2}
$$
is the normalization factor of the distribution. 

Hence, we find:
$$
\mathbb{E}(T_x) = 2t_c \sqrt{\frac{\sigma}{x}}\int_{-\infty}^{0}\sqrt{|\xi|} \frac{1}{\sqrt{2\pi}}e^{-\frac{\xi^2}{2}} = \frac{2^{1/4}}{\sqrt{\pi}}\Gamma(3/4)\  t_c\sqrt{\frac{\sigma}{x}} .
$$

As the mean value of the stochastic variable $T_x^2$ is given by
$$
\mathbb{E}(T_x^2) = \frac{1}{\mathcal{N}}\int_{-\infty}^{0}T_x(\xi)^2 \frac{1}{\sqrt{2\pi}}e^{-\frac{\xi^2}{2}} ,
$$
we obtain
$$
\mathbb{E}(T_x^2) = 2t_c^2 \frac{\sigma}{x}\int_{-\infty}^{0}|\xi| \frac{1}{\sqrt{2\pi}}e^{-\frac{\xi^2}{2}} = \sqrt{\frac{2}{\pi}}\ \frac{\sigma}{x}t_c^2.
$$

\section*{Uncertainty for the energy of a free-falling particle}

The Hamiltonian $\hat{H}$ for a free-falling quantum particle is given by:
$$
    \hat{H} = \frac{\hat{p}^2}{2m}-mg\hat{x} ,
$$
where $\hat{p} = -i\hbar\frac{\partial}{\partial x}$ is the momentum operator and $\hat{x}$ is the position operator.  
Consider the initial Gaussian wavepacket:
\begin{equation}\label{Eq:InitialWavePacket}
    \psi_0(x) = \frac{1}{(2\pi\sigma^2)^{1/4}}e^{-\frac{x^2}{4\sigma^2}}\ ,
\end{equation}
where the mean value of the particle's momentum is zero $\langle \hat{p}\rangle = 0$, and where $\sigma$ is the standard deviation of the position operator $\sqrt{\langle \hat{x}^2\rangle-\langle \hat{x}\rangle^2}=\sigma$, with $\langle \hat{x}\rangle = 0$. 
The initial Gaussian wavepacket can also be expressed in the momentum basis through the Fourier transform of \eqref{Eq:InitialWavePacket}
$$
    \psi_0(p) = \frac{1}{(2\pi \sigma_p^2)^{1/4}}e^{-\frac{p^2}{4\sigma_p^2} }\ ,
$$
where the standard deviation of the momentum operator is $\sqrt{\langle \hat{p}^2 \rangle - \langle \hat{p} \rangle^2} = \sigma_p = \frac{\hbar}{2\sigma}$. 

We may calculate the mean value $\langle \hat{H} \rangle$  of the energy operator as:
\begin{equation}\label{Eq:MeanEnergy}
\langle \hat{H} \rangle  =  \langle \frac{\hat{p}^2}{2m} \rangle - \langle mgx \rangle = \langle \frac{\hat{p}^2}{2m} \rangle + 0 
= \frac{\sigma_p^2}{2m} =  \frac{\hbar^2}{8m\sigma^2} .
\end{equation}
We also calculate the mean squared value of the energy operator as:
\begin{equation}\label{Eq:MeanEnergySquared}
\langle \hat{H}^2 \rangle  = \langle \frac{\hat{p}^4}{4m^2} \rangle + \langle m^2g^2\hat{x}^2 \rangle - mg\langle (\hat{p}^2\hat{x}+\hat{x}\hat{p}^2 )\rangle = \frac{3}{4m^2}\sigma_p^4 + m^2g^2\sigma^2 = \frac{3\hbar^4}{64m^2\sigma^4} + m^2g^2\sigma^2  ,
\end{equation}
where we used $\langle \hat{x}^2\rangle = \sigma^2$, $\langle \hat{p}^4\rangle = 3\sigma_p^4$, and $\langle (\hat{p}^2\hat{x}+\hat{x}\hat{p}^2 )\rangle =0$, as can be seen from:
\begin{align*}
\langle (\hat{p}^2\hat{x}+\hat{x}\hat{p}^2 )\rangle &= \langle (2\hat{x}\hat{p}^2-2i\hbar \hat{p})\rangle \\
&= \langle 2\hat{x}\hat{p}^2)\rangle +0  \\
&= -2\hbar^2\int_{-\infty}^{+\infty} dx\ x \psi_0(x)^\ast\frac{\partial^2}{\partial x^2}\psi_0(x) \\
&= -2\hbar^2\int_{-\infty}^{+\infty} dx\ x \psi_0(x)^\ast\frac{\partial}{\partial x}\left(-\frac{x}{2\sigma^2}\psi_0(x)\right) \\
&= -2\hbar^2\int_{-\infty}^{+\infty} dx\ x \psi_0(x)^\ast\left(-\frac{1}{2\sigma^2}+\frac{x^2}{4\sigma^4}\right)\psi_0(x) \\
&= -2\hbar^2\int_{-\infty}^{+\infty} dx\ \left(-\frac{x}{2\sigma^2}+\frac{x^3}{4\sigma^4}\right)|\psi_0(x)|^2 \\
&= -2\hbar^2\left(-\frac{\langle x\rangle}{2\sigma^2}+\frac{\langle x^3\rangle}{4\sigma^4}\right),
\end{align*}
and the announced result follows from since $\langle x\rangle = 0$ and $\langle x^3\rangle = 0$. Combining \eqref{Eq:MeanEnergy} and \eqref{Eq:MeanEnergySquared}, we find that the standard deviation of the energy operator is given as:
\begin{equation}
    \Delta E^2 \equiv \langle \hat{H}^2\rangle - \langle \hat{H}\rangle^2 = \frac{3\hbar^4}{64m^2\sigma^4} + m^2g^2\sigma^2 - \frac{\hbar^4}{64m^2\sigma^4} = \frac{\hbar^4}{32m^2\sigma^4} + m^2g^2\sigma^2 =  m^2g^2\sigma^2\left(1+\frac{\hbar^4}{32m^4g^2\sigma^6}\right) ,
\end{equation}
from which we finally obtain:
\begin{equation}
    \Delta E = mg\sigma\sqrt{1+\frac{\hbar^4}{32m^4g^2\sigma^6}} .
\end{equation}

\end{document}